\documentclass{article}

\usepackage{amsthm,amsmath,amsfonts}
\usepackage[authoryear]{natbib}
\usepackage{algorithm}
\usepackage[noend]{algpseudocode}
\usepackage{graphicx}
\usepackage{url}
\usepackage{natbib}
\usepackage{authblk}

\DeclareMathOperator*{\argmin}{arg\,min}
\newcommand{\sgn}{\operatorname{sgn}}

\newcommand\blfootnote[1]{%
  \begingroup
  \renewcommand\thefootnote{}\footnote{#1}%
  \addtocounter{footnote}{-1}%
  \endgroup
}

\title{High-dimensional regression\\ over   disease subgroups}

\author[1]{Frank Dondelinger}
\author[2]{Sach Mukherjee}
\author[ ]{The Alzheimer's Disease Neuroimaging Initiative*}
\affil[1]{Lancaster Medical School, Lancaster University}
\affil[2]{DZNE, Bonn}

\begin{document}

\maketitle

\bigskip

\begin{abstract}
We consider high-dimensional regression over subgroups of observations.
Our work is motivated by biomedical problems, where disease subtypes, for example, may differ with respect to underlying regression models, but sample sizes at the subgroup-level may be limited.
We focus on the case in which subgroup-specific models
may be expected to be similar but not necessarily identical.
Our approach is to treat subgroups as related problem instances and jointly estimate subgroup-specific regression coefficients.
This is done in a penalized framework, combining an  $\ell_1$ term  with an additional term that penalizes differences between subgroup-specific coefficients. This gives solutions that are globally sparse but that allow information-sharing between the subgroups.
We present  algorithms for estimation and empirical results on simulated data and using Alzheimer's disease, amyotrophic lateral sclerosis
 and cancer datasets.
These examples demonstrate the gains our approach can offer in terms of  prediction and the ability to estimate subgroup-specific sparsity patterns.

\end{abstract}



\blfootnote{*Data used  in  preparation  of  this  article  were  obtained  from  the  Alzheimer’s  Disease Neuroimaging  Initiative  (ADNI)  database  (adni.loni.usc.edu). As  such,  the  investigators within the ADNI contributed to the design and implementation of ADNI and/or provided data but  did  not  participate in analysis or writing of this report. A complete listing of ADNI investigators can be found at: \url{http://adni.loni.usc.edu/wp-content/uploads/how_to_apply/ADNI_Acknowledgement_List.pdf}}

\section{Introduction}
High-dimensional regression has been  well studied in the case where all samples can reasonably be expected to follow the same  model. However, in several current and emerging applications, observations span multiple subgroups that  may not be identical with respect to the underlying regression models. Examples abound, from disease subtypes in biomedicine to customer subsets in business applications. We are specifically motivated by biomedical problems, where  sets of samples, such as disease subtypes, although related, may differ with respect to underlying biology and therefore have different relationships between covariates and a response of interest.

Thus, we focus on high-dimensional regression in group-structured settings. In particular, we consider linear regression in a commonly-encountered scenario in which the same set of $p$ covariates or predictors is available in each of $K$ subgroups.  That is, we consider subgroup-specific linear regression problems indexed by $k$, each with
subgroup-specific sample size $n_k$, a response vector $y_k$ of length $n_k$, a $n_k \times p$ feature matrix $X_k$ and a $p$-vector $\beta_k$ of regression coefficients. The problem we address is estimating  the regression coefficients $\beta_1 \ldots \beta_K$.

We propose an approach to jointly estimate the regression coefficients that induces global sparsity and encourages similarity between subgroup-specific coefficients.
We consider the following penalized formulation and its variants
\begin{eqnarray*}
\hat{B} & = & \argmin_{B=[\beta_1 \ldots \beta_k]} \sum_{k=1}^K \left\{ \|y_k - X_k \beta_k\|_2^2 + \lambda \| \beta_k \|_1  + \gamma \sum_{k' > k} \tau_{k, k'} \| \beta_k - \beta_{k'} \|_2^2 \right\}
\end{eqnarray*}

\noindent
where, $B=[\beta_1 \ldots \beta_K]$ is a $p \times K$ matrix that collects together all the regression coefficients, $\| \cdot \|_q$ denotes the $\ell_q$ norm of its argument and $\lambda, \gamma, \tau$ are tuning parameters. The last term is a fusion-type penalty between subgroups;
note that the difference is taken between entire vectors of subgroup-specific coefficients.
An $\ell_2$ fusion penalty is shown above, although other penalties may be used; in this manuscript, we also consider an  $\ell_1$  variant.
The parameters $\tau_{k,k'}$  allow for the possibility of controlling the extent to which
similarity is encouraged for specific pairs of subgroups.

Our proposal shares similarities with both the group lasso \citep{yuan2006} and the fused lasso \citep{tibshirani_05}
but differs from both in important ways.
 In contrast to the group lasso, we consider subgroups of samples or observations rather than groups of coefficients and in contrast to the fused lasso, we consider fusion of entire (subgroup-specific) coefficient vectors, rather than successive coefficients under a pre-defined ordering. \citet{obozinski2010} showed how the group lasso could be used in subgroup-structured settings, essentially by considering the global problem  and defining groups (in the group lasso sense) corresponding to the same covariate across all subgroups. This means that each covariate tends either to be included in all subgroup-specific models or none. In contrast, our approach allows subgroups to have different sparsity patterns, whilst pulling subgroup-specific coefficients together and inducing  global sparsity. Our work is also similar in spirit to recent work  concerning joint estimation of graphical models over multiple problem instances \citep{danaher2014,oates2014,oates2015}.

We are  motivated by emerging problems in biomedical research and specifically in personalized medicine.
High-dimensional regression problems are now becoming common in this area, with several high-dimensional data types already in mainstream use.
In the personalized  medicine setting, samples usually correspond to individuals and the subgroups $k$  to e.g.  diseases or disease subtypes. It is increasingly clear that many disease subtypes differ in their biology \citep[see e.g.][]{weinstein2013,akbani2014}, suggesting that relationships between covariates and responses of interest may differ between them. However, sample sizes tend to be limited, especially at the subgroup level, posing
 problems for the subgroup-wise strategy of solving each problem separately. On the other hand, pooling all the data together into a single regression problem may lead to severe mis-specification if the underlying subgroup-specific models do indeed differ.

These issues  may lead to  losses in terms of  predictive ability and perhaps just as important in the ability to efficiently estimate subgroup-specific influences  that may themselves be of  interest.
In contrast to simple pooling, our approach allows subgroups to have different sparsity patterns and regression coefficients, but in contrast to the subgroup-wise approach it  takes advantage of similarities between  subgroups.

We show empirical results in the context of two neurodegenerative diseases   - Alzheimer's disease and amyotrophic lateral sclerosis (ALS) - and cancer (see below for full details of the applications and data).
The responses concern disease progression in Alzheimer's and ALS and therapeutic response in cancer cell lines.
In the Alzheimer's and ALS examples, subgroups are based on clinical factors, while in the cancer data they are based on the tissue type of the cell lines.

Across the three examples, data types include genetic, clinical and transcriptomic variables.
We find that our approach can improve performance relative to pooling or subgroup-wise analysis. Importantly, in cases where pooling or subgroup-wise analyses do well (perhaps reflecting a lack of subgroup structure or insufficient similarity respectively) our approach remains competitive. This gives assurance that penalization is indeed able to share information appropriately in real-world examples.
We emphasize that the goal of the empirical  analyses we present is not to
give the best predictions possible
in these applications,
but rather to explore joint estimation in group-structured  biomedical problems.

\section{Methods}

\subsection{Notation}
Each subgroup $k \in \{1 \ldots K \}$ has the same set of $p$ covariates, but  subgroup-specific sample size $n_k$. Total sample size is $n=\sum_{k =1}^K n_k$. For subgroup $k$, $X_k$ is the  $n_k \times p$ feature matrix and $y_k$ the corresponding $n_k \times 1$ vector of observed responses. Subgroup-specific regression coefficients are $\beta_k \in \mathbb{R}^p$. Where convenient we collect all regression coefficients together in a $p \times K$ matrix $B = [\beta_1 \ldots \beta_K]$ and accordingly we use $\beta_{j,k}$ to denote the coefficient for covariate $j$ in subgroup $k$.

\subsection{Model Formulation}

We seek to jointly estimate the regression coefficients $B = [\beta_1 \ldots \beta_K]$ whilst ensuring global sparsity and encouraging agreement between subgroup-specific coefficients. We propose the criterion
\begin{equation}
\label{eq:model_l2_fusion}
\hat{B}  =  \argmin_{B= [\beta_1 \ldots \beta_K]} \sum_{k=1}^K \left\{ \|y_k - X_k \beta_k\|_2^2 + \lambda \| \beta_k \|_1  + \gamma \sum_{k' > k} \tau_{k, k'} \| \beta_k - \beta_{k'} \|_2^2 \right\}
\end{equation}

\noindent
and a variant with an $\ell_1$ norm in the last term
\begin{equation}
\label{eq:model_l1_fusion}
\hat{B}  =  \argmin_{B} \sum_{k=1}^K \left\{ \|y_k - X_k \beta_k\|_2^2 + \lambda \| \beta_k \|_1  +\gamma \sum_{k' > k} \tau_{k, k'} \| \beta_k - \beta_{k'} \|_1 \right\} .
\end{equation}

\noindent
Here, $\lambda, \gamma, \tau$ are tuning parameters.
The role of the last term is to encourage similarity between subgroup-specific regression coefficients.
The special case $K=1$ recovers the classical lasso (applied to all data pooled together).
The tuning parameters $\tau_{k,k'}$  give the possibility of controlling the extent of fusion between specific subgroups.
By default all $\tau$'s are set to unity (``unweighted fusion"), but they can also be set to specific values as discussed below (``weighted fusion"). In the above formulation, we assume that $y_k$ and $X_k$ have been standardized (at the subgroup level) so that no intercept terms are required. Note that the regularization parameters $\lambda, \gamma$ are the same across subgroups.

The difference between the two variants is that the first, $\ell_2$ fusion encourages similarity between subgroup-specific coefficients, while the second $\ell_1$ version allows for exact equality.
The $\ell_2$ formulation has the computational advantage that the fusion part of the objective function becomes continuously differentiable, and the estimate of the objective function at each step can be obtained by soft-thresholding, analogously to coordinate descent for regular lasso problems. In the $\ell_1$ formulation on the other hand, the fusion constraint is only piece-wise continuously differentiable, leading to a more difficult optimisation problem (see  below).

\subsection{Comparison with group and fused lasso}
Our formulation resembles the group lasso and fused lasso, but differs from both in important ways.
The original group lasso \citep{yuan2006} was designed to consider groups of covariates within a single regression problem. Let $X$ be the feature matrix and $y$ the vector of responses in a standard regression problem. Letting $l \in \{1 \ldots L \}$ index groups of covariates, the group lasso criterion is
\begin{eqnarray}
\hat{\beta} & = & \argmin_{\beta} \| y - \sum_{l=1}^L X^{(l)} \beta^{(l)} \|_2^2 + \sum_{l=1}^L \lambda_l  \|\beta^{(l)} \|_2
\end{eqnarray}

\noindent
where $X^{(l)}$ is the submatrix of $X$ corresponding to the covariates in group $l$, $\beta^{(l)}$ the corresponding regression coefficients and $\lambda_l$'s tuning parameters. The penalty tends to include or exclude all members of a group from the model, i.e. all coefficients in a group may be set to zero giving groupwise sparsity.

In our setting, the subgroups are subsets of samples rather than covariates. Nevertheless,
as shown in \citet{obozinski2010}, one could use a group lasso-like criterion for estimation in the multiple subgroup setting by forming groups $l$ each comprising all the coefficients for a single covariate $j \in \{ 1 \ldots p \}$ across all $K$ regression problems. This encourages covariates to either be included in all the subgroup-specific models or none.

The fused lasso \citep{tibshirani_05} is also aimed at a single regression problem, but assumes that the covariates can be ordered in such a way that successive coefficients may be expected to be similar. This leads to the following criterion
\begin{equation}
\label{eq:sim_fused_1d}
 \hat{\beta} = \argmin_{\beta} \|y - X \beta \|^2_2 + \lambda \|\beta \|_1 + \gamma \sum_{i=1}^{p-1} \|\beta_i - \beta_{i+1} \|_1
\end{equation}

\noindent
where $\lambda,\gamma$ are tuning parameters and
we have assumed that the covariates are in a suitable order. The final term encourages similarity between successive coefficients.
Efficient solutions for various classes of this problem exist \citep[e.g.][]{hoefling_10,liu_10,ye_11}.

Our approach shares the use of a fusion-type penalty, but focuses on a different problem, namely that of jointly estimating regression coefficients across multiple, potentially non-identical, problems. Accordingly, our  penalty encourages agreement between entire coefficient vectors from different subgroups and does not require any ordering of covariates.


%
%

%
%

\subsection{Setting the tuning parameters $\tau$}
For weighted fusion, the parameters $\tau_{k,k'}$ could be set by cross-validation but this may be  onerous in practice. As an alternative we consider setting $\tau_{k,k'}$ using a distance function $d(k, k')$ based on the covariates. The idea is to allow more fusion between subgroups that are similar with respect to $d$, while allowing the $\tau_{k,k'}$  to  be set in advance of estimation proper. However, this assumes that similarity in the covariates
reasonably reflects similarity between the underlying regression coefficients, which may or may not be the case in specific applications.

We consider two variants.
The first sets
$d(k,k') = \| \mu_k - \mu_{k'} \|_2$
where $\mu_k, \mu_{k'}$ are the sample means of the covariates in the subgroups $k,k'$ respectively
 (we assume the data have been standardized).
The second approach additionally  takes the covariance structure into account by using the symmetrised Kullback-Leibler (KL) divergence, i.e. $d(k,k') = \frac{1}{2}(\mathrm{KL}(\hat{p}_k \| \hat{p}_{k'}) + \mathrm{KL}(\hat{p}_{k'} \| \hat{p}_k))$, where
$\hat{p}_k, \hat{p}_{k'}$ are estimated marginal  distributions over the covariates in the subgroups $k,k'$ respectively and $\mathrm{KL}(p \| q)$ is the KL-divergence between distributions $p$ and $q$. In practice,  this requires  simplifying distributional assumptions. Below we use multivariate Normal models for this purpose, with the graphical lasso \citep{friedman_08} used to estimate the $\Sigma_k$'s.
For both approaches, we set $\tau_{k,k'} =  1 - d(k,k')/d_{max}$, with $d_{max}$ the largest distance between any pair of groups $k$,$k'$ (this scales $\tau$ to the unit interval).

\subsection{Optimisation}

We describe a coordinate descent approach for optimising equation (\ref{eq:model_l2_fusion}). While it is possible to derive a block coordinate descent approach for equation (\ref{eq:model_l1_fusion}) \citep[e.g. following][]{friedman_07}, this is generally inefficient for the high-dimensional problems that we consider. Instead, we will describe an optimization procedure based on a proximal gradient approximation derived in \cite{chen_10}.

\subsubsection{Coordinate Descent for $\ell_2$ Fusion}
\label{sec:optimisation_l2}

The $\ell_2$ fusion penalty is continuously differentiable and we can obtain the optimal value for $\hat{\beta}_{j,k}$ in equation (\ref{eq:model_l2_fusion}) at each step by first calculating  optimal values without the lasso penalty:

\begin{equation}
 \label{eq:l2_update}
 \hat{\beta}^*_{k,j} = \frac{x_{j,k}^T (y_k - X_{-j,k}\beta_{-j,k}) + 2 \gamma \sum_{k' \neq k} \tau_{k,k'} \beta_{j,k'}} {x_{j,k}^T x_{j,k} + 2 \gamma \sum_{k' \neq k}\tau_{k,k'}}
\end{equation}

Then $\hat{\beta}_{j,k}$ can be obtained by soft-thresholding on $\hat{\beta}^*_{k,j}$. The procedure is summarized in Algorithm \ref{alg:block_descent_l2}.

\bigskip
\begin{algorithm}
\caption{Block Coordinate Descent}\label{alg:block_descent_l2}
\begin{algorithmic}[1]
\Procedure{BlockDescentL2}{$n_{iter}$, $X$, $Y$, $\beta_{init}$, $\lambda$, $\gamma$, $\tau$}
\State $i \gets 0$
\State $\beta \gets \beta_{init}$
\While{not\_converged AND $i < n_{iter}$}

  \ForAll{j in 1:P}
    \State $\beta^{temp}_{j,1:K} \gets$ \Call{DescentUpdateL2}{$X$, $Y$, $\beta$, $j$, $\gamma$, $\tau$} using eq. (\ref{eq:l2_update})
    \State $\beta_{j,1:K} \gets \sgn(\beta^{temp}_{j,1:K}) * \max(\beta^{temp}_{j,1:K} - \lambda, 0)$
  \EndFor

  \State $i \gets i + 1$
\EndWhile
\EndProcedure
\end{algorithmic}
\end{algorithm}

\bigskip
While Algorithm \ref{alg:block_descent_l2} is easy to understand and implement, a naive implementation in most programming languages will be still be slow due to the need for an inner for-loop over $p$, where $p$ can be in the tens of thousands for the kinds of problems we will consider. In order to efficiently optimize $B$, we reformulate (\ref{eq:model_l2_fusion}) as a classical lasso problem and apply the \texttt{glmnet} software \citep{friedman_10}. We transform the sum in first part of the objective into matrix form $y_{flat} - X_{diag}b_{flat}$ by defining $X_{diag}$ as a block-diagonal $n \times pK$ matrix with $X_k$ along the diagonals. The vector $b_{flat}$ is  a flattened version of $B$ with stacked $\beta_k$ vectors, and similarly for $y_{flat}$. So we have:

\noindent
\begin{minipage}{.5\linewidth}
\begin{equation*}
  X_{diag}  = 
 \begin{pmatrix}
  X_1 &  &  \\
      & \ddots & \\
      &        & X_K
 \end{pmatrix}
\end{equation*}
\end{minipage}%
\begin{minipage}{.25\linewidth}
\begin{equation*}
  b_{flat} = \begin{pmatrix}
           \beta_{1} \\
           \vdots \\
           \beta_{K}
         \end{pmatrix}
\end{equation*}
\end{minipage}%
\begin{minipage}{.25\linewidth}
\begin{equation*}
  y_{flat} = \begin{pmatrix}
           y_{1} \\
           \vdots \\
           y_{K}
         \end{pmatrix} 
\end{equation*}
\end{minipage}
~

Now we move the $\ell_2$ fusion penalty into the first squared term by defining the augmented matrix $X^{aug}_{diag}$, and augmented vector $y^{aug}_{flat}$, such that

\begin{equation}
\label{eq:model_l2_fusion_glmnet}
\hat{b}_{flat}  =  \argmin_{b_{flat}} \|y^{aug}_{flat} - X^{aug}_{diag} b_{flat}\|_2^2 + \lambda \| b_{flat} \|_1 
\end{equation}

\noindent
where

\noindent
\begin{minipage}{.5\linewidth}
\begin{equation*}
  X^{aug}_{diag}  = 
 \begin{pmatrix}
  X_{diag} \\
  \Gamma
 \end{pmatrix}
\end{equation*}
\end{minipage}%
\begin{minipage}{.5\linewidth}
\begin{equation*}
  y^{aug}_{flat} = \begin{pmatrix}
           y_{flat} \\
           \vec{0}
         \end{pmatrix}
\end{equation*}
\end{minipage}

\bigskip
\noindent
with $\Gamma$ a $pK(K-1)/2 \times pk$ matrix encoding the pair-wise fusion constraints, and $\vec{0}$ a $pK(K-1)/2 \times 1$ vector of zeros. Each block $\Gamma_{k,k'}$, $k,k' \in [1,K], k<k'$ of $p$ rows of $\Gamma$ corresponds to the fusion constraint between two coefficient vectors $\beta_k$ and $\beta_{k'}$, with:

\begin{equation}
\Gamma_{k,k'}(l,m) =
  \begin{cases}
    \gamma \tau_{k, k'}       & \quad \text{if } l=p(k-1)+m  \\
    -\gamma \tau_{k, k'}  & \quad \text{if } l=p(k'-1)+m\\
    0 & \quad \text{otherwise.}
  \end{cases}
\end{equation}

We can see that (\ref{eq:model_l2_fusion_glmnet}) is  a classical lasso problem, to which \texttt{glmnet} can be directly applied.

\subsubsection{Proximal-Gradient Approach for Fused L1 Penalty}

Optimising equation (\ref{eq:model_l1_fusion}) by block gradient descent, while possible, is highly inefficient due to having to deal with the discontinuities in the objective function space. In \cite{chen_10}, the authors describe a proximal relaxation of this problem that introduces additional smoothing to turn the objective function $f_{L1}(B)$ into a continuously differentiable function $f^{\mu}_{L2}(B)$. Chen et al. deal with the multi-task regression setting (with common $X$ for each task); it is straightforward to adapt their procedure for the subgroup regression setting with different $X_k$ per subgroup. 

It is notationally convenient to first introduce a graph formulation of the fusion penalties. We will think of the fusion constraints in terms of an undirected graph $G=(V,E)$ with vertex set $V = \{ 1 \ldots K \}$ corresponding to the subgroups and edges between all vertices.
Then the $\ell_1$ penalised objective function can be written as:
\begin{equation}
 \label{eq:sim_fused_l1}
 f_{L1}(B) = \sum_k \{\|y_k - X_k \beta_k\|^2_2\} + \| BC \|_1
\end{equation}

\noindent
where the last term includes  both sparsity and fusion penalties, via the matrix $C=(\lambda I_K, \gamma H)$, with $I_K$ the identity matrix of size $K$, $C$ a $K \times |E|$ matrix ($|E| = {\binom{K}{2}}$ in this case) and $H_{k,e}=\tau_{k,l}$ if if $e=(m,l)$ and $k=l$, $H_{k,e}=-\tau_{m,k}$ if $e=(m,l)$ and $k=l$. Note that unlike in \cite{chen_10}, we require the explicit sum over $k$ in the objective to account for different sample sizes $N_k$ in different groups\footnote{It would be possible to reformulate the first part of the objective in matrix form $Y_{diag} - X_{diag}B_{diag}$ by defining $X_{diag}$ as a block-diagonal matrix as in Section \ref{sec:optimisation_l2}, defining $B_{diag}$ as a $pK \times K$ matrix with $\beta_k$ along the diagonals and similarly $y_{diag}$ as an $n \times K$ matrix with $y_k$ along the diagonals; however, this formulation is neither practical nor intuitive, and the gain in notational simplicity is negligible.}.

The graph formulation allows for zero edges by setting $\tau_{k,k'}$ to zero. We have implicitly assumed in the formulation of (\ref{eq:model_l2_fusion}) and (\ref{eq:model_l1_fusion}) that the relationship between subgroups is represented by an undirected graph. However, (\ref{eq:sim_fused_l1}) is completely general, and it would be straightforward to incorporate a directed graph in our model. We have not pursued this avenue here, as there is no reason to suspect directionality in the subgroup relationships for the applications we consider below, and including directionality would double the number of tuning parameters $\tau_{k,k'}$ that need to be considered.

Following \cite{chen_10}, we can introduce an auxiliary matrix $A \in \mathcal{Q}=\{A'| ~ \|A'\|_{\infty} \leq 1, A' \in \mathcal{R}^{p \times (K + |E|)}\}$. Because of duality between $\ell_1$ and $\ell_\infty$, we can write $\|BC\|_1 = \max_{\|A\|_\infty \leq 1} \langle A, BC \rangle$. A smooth approximation of $\|BC\|_1$ is then obtained by writing:

\begin{equation}
\label{eq:smooth_approx}
f_\mu(B) = \max_{\|A\|_\infty \leq 1} \langle A, BC \rangle - \mu d(A)
\end{equation}  

where $\mu$ is a positive smoothness parameter, and $d(A) \equiv \frac{1}{2}\|A\|^2_F$, with $\|\cdot\|_F$ the Frobenius norm. They show that for a desired accuracy $\epsilon$, we need to set $\mu=\frac{\epsilon}{p(K+|E|)}$. Theorem 1 in \cite{chen_10} gives the gradient of $f_\mu(B)$ as $\Delta f_\mu(B) = A^*C^T$, where $A^*$ is the optimal solution of (\ref{eq:smooth_approx}). Replacing $\|BC\|_1$ by $f_\mu(B)$ in equation (\ref{eq:sim_fused_l1}), we obtain

\begin{equation}
\label{eq:proximal_fused_l1}
 \tilde{f}_{L1}(B) = \sum_k \{\|y_k - X_k \beta_k\|^2_2\} + f_\mu(B)
\end{equation}

\noindent
which is now continuously differentiable with gradient

\begin{equation}
\label{eq:proximal_fused_l1_grad}
 \Delta \tilde{f}_{L1}(B) = \sum_k \{X_k^T(X_K\beta_k-y_k)\} + f_\mu(B) \, .
\end{equation}

Chen et al. further show that $A^*=S(BC/\mu)$ where function S truncates each entry of $A^*$ to the range [-1,1] to ensure that $A^* \in \mathcal{Q}$. An upper bound $L_U$ of the Lipschitz constant L can be derived as:

\begin{equation}
\label{eq:lipschitz_bound}
 L_U = \max_k (\lambda_{max}(X_k^TX_k)) + \frac{\lambda^2+2*\gamma^2\max_{k\in V}d_k}{\mu}
\end{equation}

\noindent
where $\lambda_{max}(M)$ is the largest eigenvalue of $M$ and $d_k = \sum_{k'}^K \tau_{k,k'}$.

With the derivation of the gradient in (\ref{eq:proximal_fused_l1_grad}) and the Lipschitz bound in (\ref{eq:lipschitz_bound}), we can now apply Nesterov's method \citep{nesterov_05} for optimizing (\ref{eq:proximal_fused_l1}). The procedure is summarized in Algorithm \ref{alg:proximal}. For more details on the proximal approach see \cite{chen_10}.

\begin{algorithm}
\caption{Proximal Gradient Optimization}\label{alg:proximal}
\begin{algorithmic}[1]
\Procedure{Proximal}{$n_{iter}$, $X$, $Y$, $B_{init}$, $\lambda$, $\gamma$, $\tau$, $L_U$, $\mu$}
\State $i \gets 0$
\State $W^0 \gets B_{init}$
\While{not\_converged AND $i < n_{iter}$}
  \State Compute $\Delta \tilde{f}_{L1}(W^i)$ according to (\ref{eq:proximal_fused_l1_grad}).
  \State $B^i \gets W^i - \frac{1}{L_U}\Delta \tilde{f}_{L1}(W^i)$
  \State $Z^i \gets -\frac{1}{L_U}\sum_{j=0}^i \frac{j+1}{2}\Delta \tilde{f}_{L1}(W^f)$
  \State $W^{i+1} \gets \frac{i+1}{i+3}B^i + \frac{2}{i+3}Z^i$
  \State $i \gets i + 1$
\EndWhile
\EndProcedure
\end{algorithmic}
\end{algorithm}

\section{Simulation Study}

To test the performance of the proposed approach, we simulated  data from a model
based on characteristics of a recent cancer dataset, the Cancer Cell Line Encyclopedia
\citep[CCLE; ][]{barretina2012}.
We treat cancer types as subgroups. To simulate data, we first estimated means and covariance matrices $\mu_k, \Sigma_k$ for each of $K=9$ subgroups (the eight cancer types with the latest sample sizes in CCLE plus a ninth for all other cancer types; covariances were estimated using the graphical lasso). For each group $k$, we then sampled covariates from the multivariate normal   $\mathcal{N}(\mu_k, \Sigma_k)$. For a given total sample size $n$, subgroup sizes were consistent with those in the original data.
We used a random subset of 200 gene expression levels (i.e. the dimensionality was fixed at
$p=200$). This parametric approach allowed us to vary sample sizes freely, including the case of total $n$ larger than in the original dataset. The set-up is intended to roughly reflect  the correlation structure of the covariates, but we do not expect it to capture all aspects of the real data.


We are interested in the situation in which it may be useful to share information between subgroups. But we are also interested in investigating  performance in settings that do not agree with our model formulation (the extreme cases being where  subgroups are either entirely dissimilar or identical).
Let $V = \{1 \ldots K \}$ be the set of subgroup indices (here, $K=9$). We set regression coefficients to be identical in a subset $V_0 \subseteq V$ of the subgroups, such that the size $K_0 = |V_0|$ of the subset governs the extent to which fusion could be useful.
Specifically, if $K_0=K$, all subgroups have the same regression coefficients (i.e. favoring a pooled analysis using a single regression model) and at the other extreme if $K_0=1$ all groups have differently drawn  coefficients. Intermediate values of $K_0$ give differing levels of similarity.



For a given value $K_0$, we defined membership of $V_0$ by considering the differences between the subgroup-specific models for the covariates. Specifically, we choose the $K_0$ groups that minimized the sum of symmetrised KL divergences between subgroup-specific models.
%
%
%
A coefficient vector was then drawn separately for each subgroup $k \notin V_0$ and one, shared coefficient vector drawn for all  $k \in V_0$.
Each draw was done as follows. We first sampled a binary vector $b$ of length $p$ from a Bernoulli, i.e.
$b_i \sim \mbox{Bernoulli}(0.1)$. Then we drew $ \beta_i \sim \mathcal{N}_{trunc}(0,1)$ if $b_i = 1$
and set $\beta_i=0$ otherwise, where $\mathcal{N}_{trunc}(0,1)$ denotes a
standard Normal with the interval $(-0.1,0.1)$ excluded (this is to ensure non-zero coefficients are not very small in magnitude).
Note that in the case of $K_0=1$, all groups have separately drawn coefficients and the between-subgroup KL divergence plays no role.

We compare our approaches with pooled and subgroup-wise analyses. These are performed using classical lasso (we use the \texttt{glmnet} implementation) on respectively the whole dataset or each subgroup separately.

\begin{figure}[htbp]
  \centering
  \includegraphics[width=0.65\textwidth]{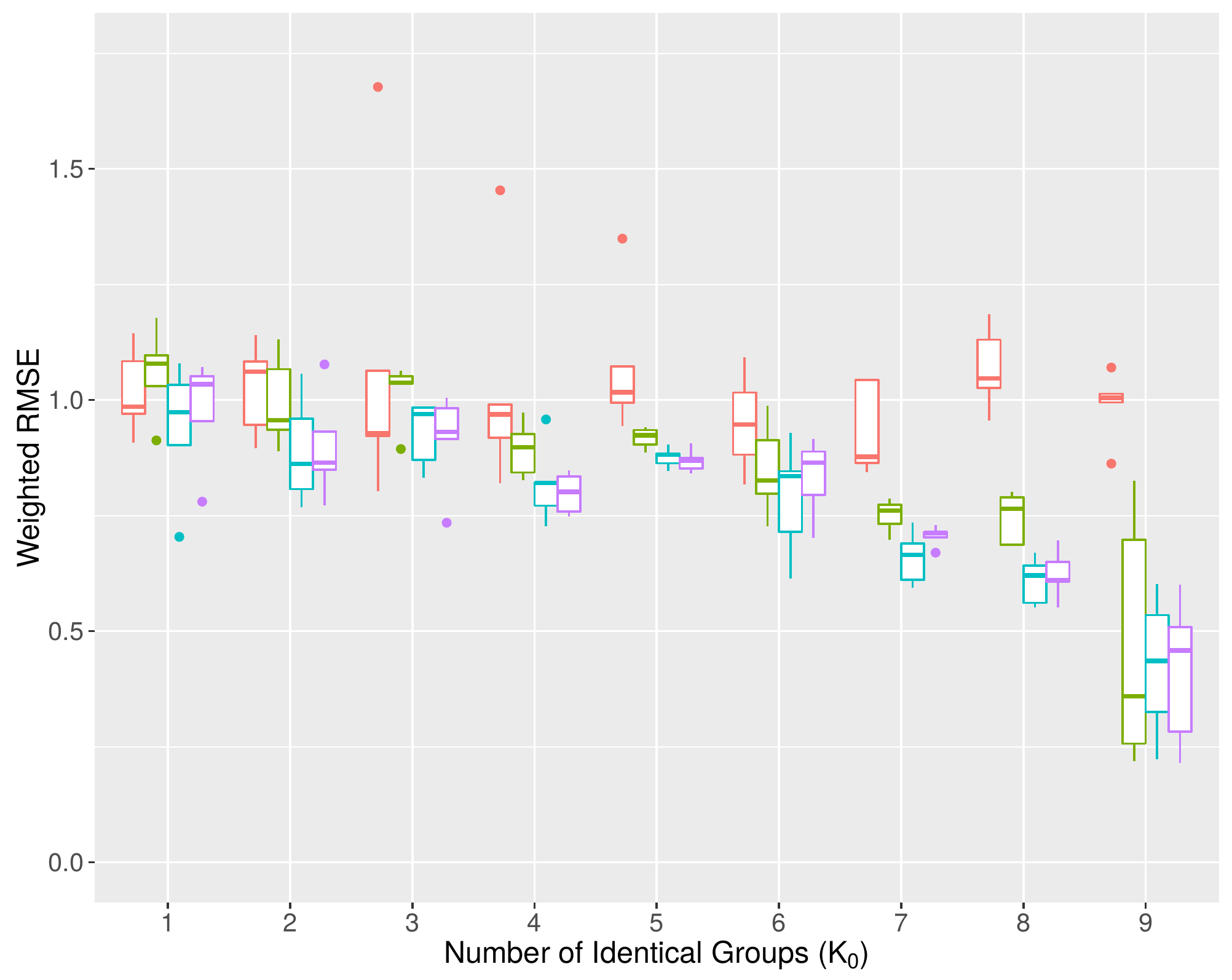}\\
  \includegraphics[width=0.65\textwidth]{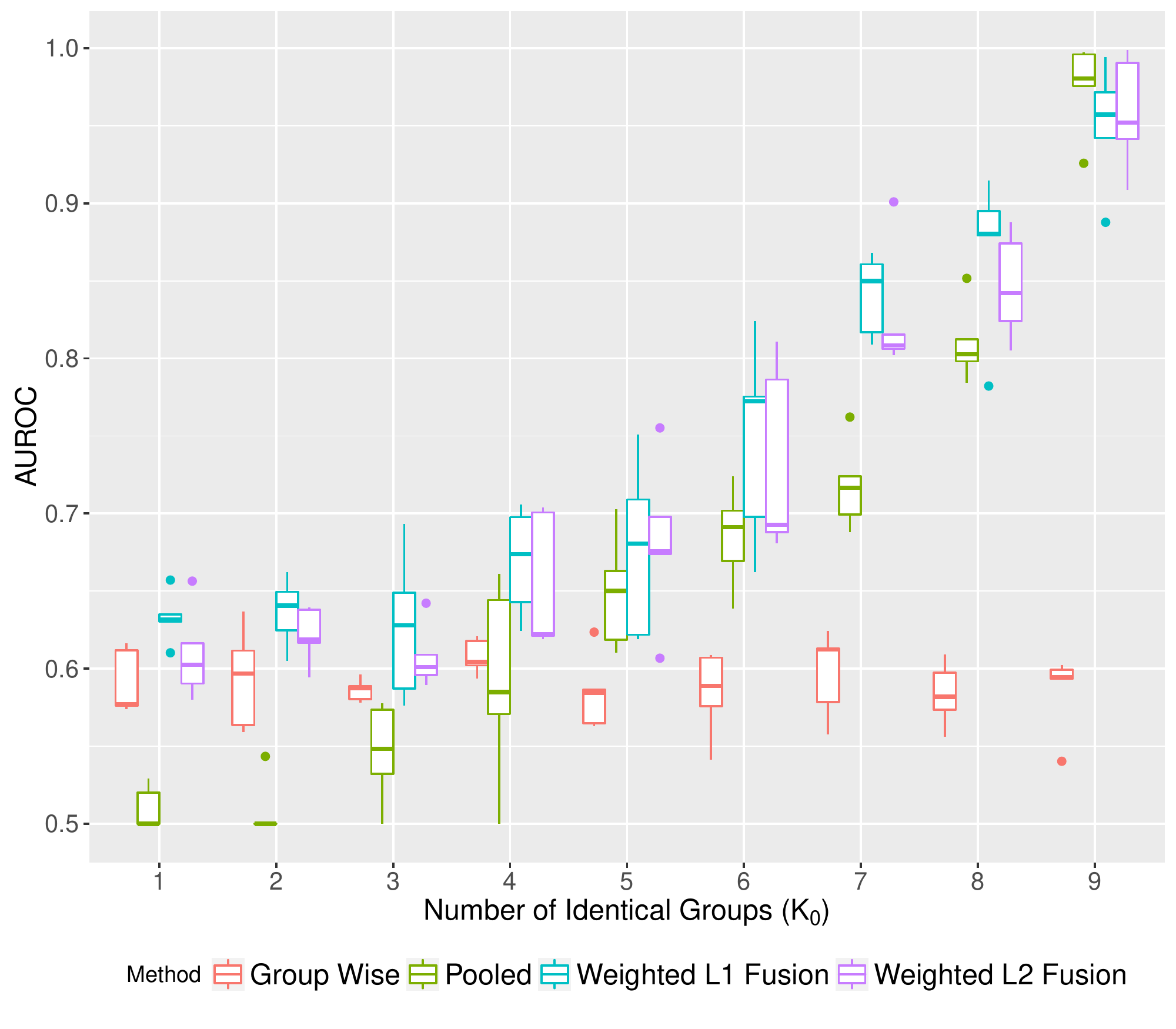}
  \caption{Simulated data, performance for varying $K_0$.
  Here, the number of subgroups is fixed at $K=9$ of which $K_0$ have shared coefficients in the underlying data-generating model (see text for details of simulation set-up).  
    A smaller $K_0$ corresponds to less similarity between underlying subgroup-specific models, with $K_0=1$ representing the case where all subgroups have separately drawn coefficients while $K_0=9$ represents an entirely homogenous model in which each subgroup has exactly the same regression coefficients.
  The total sample size is fixed at $n=250$.
Upper panel: Weighted root mean squared error (RMSE). RMSE is weighted by subgroup sizes.
Lower panel: Area under the ROC curve (AUROC; with respect to the true sets of active variables with non-zero coefficients).
}
  \label{fig:sim_vary_s}
\end{figure}

\begin{figure}[t]
  \includegraphics[width=0.45\textwidth]{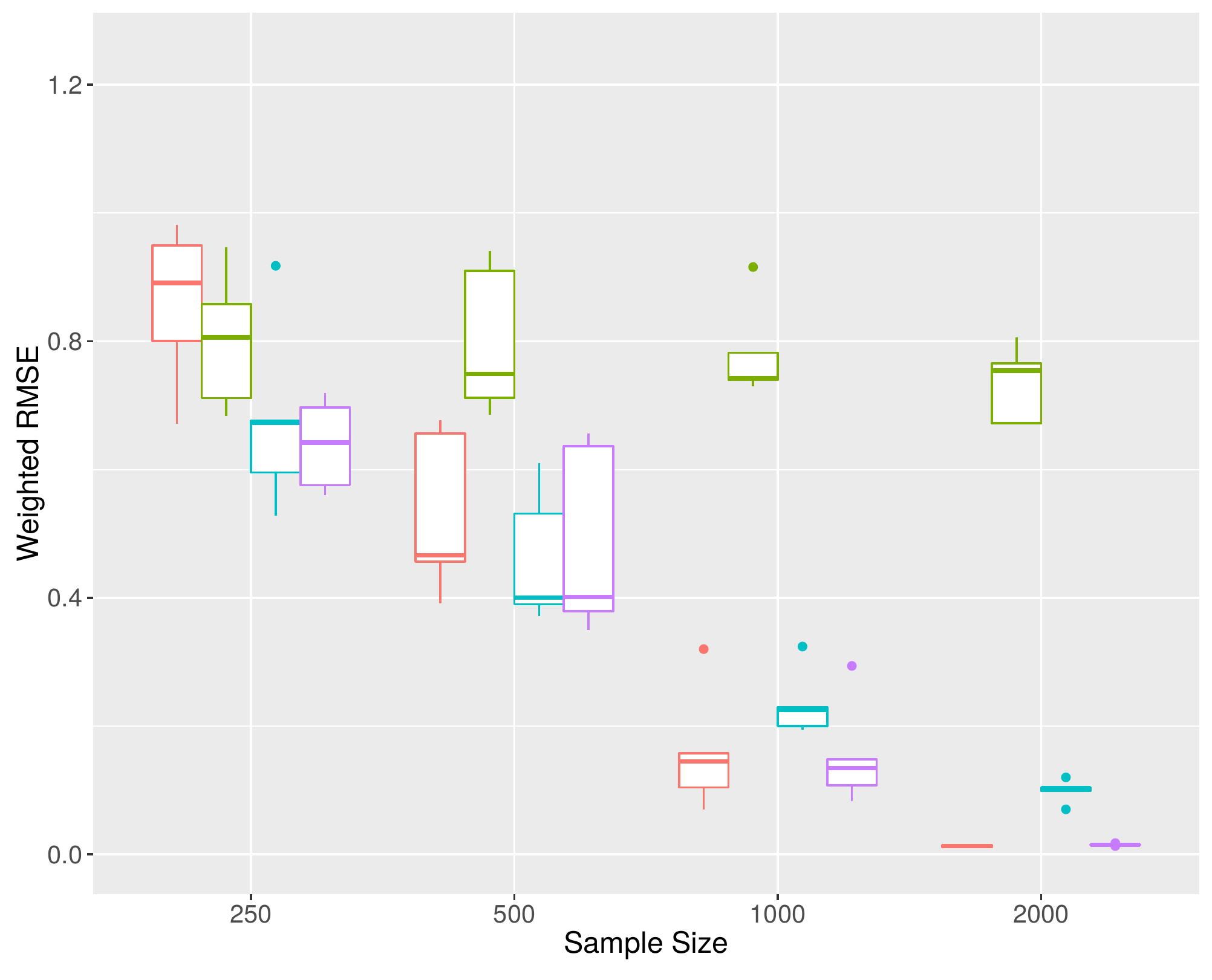}
  \includegraphics[width=0.45\textwidth]{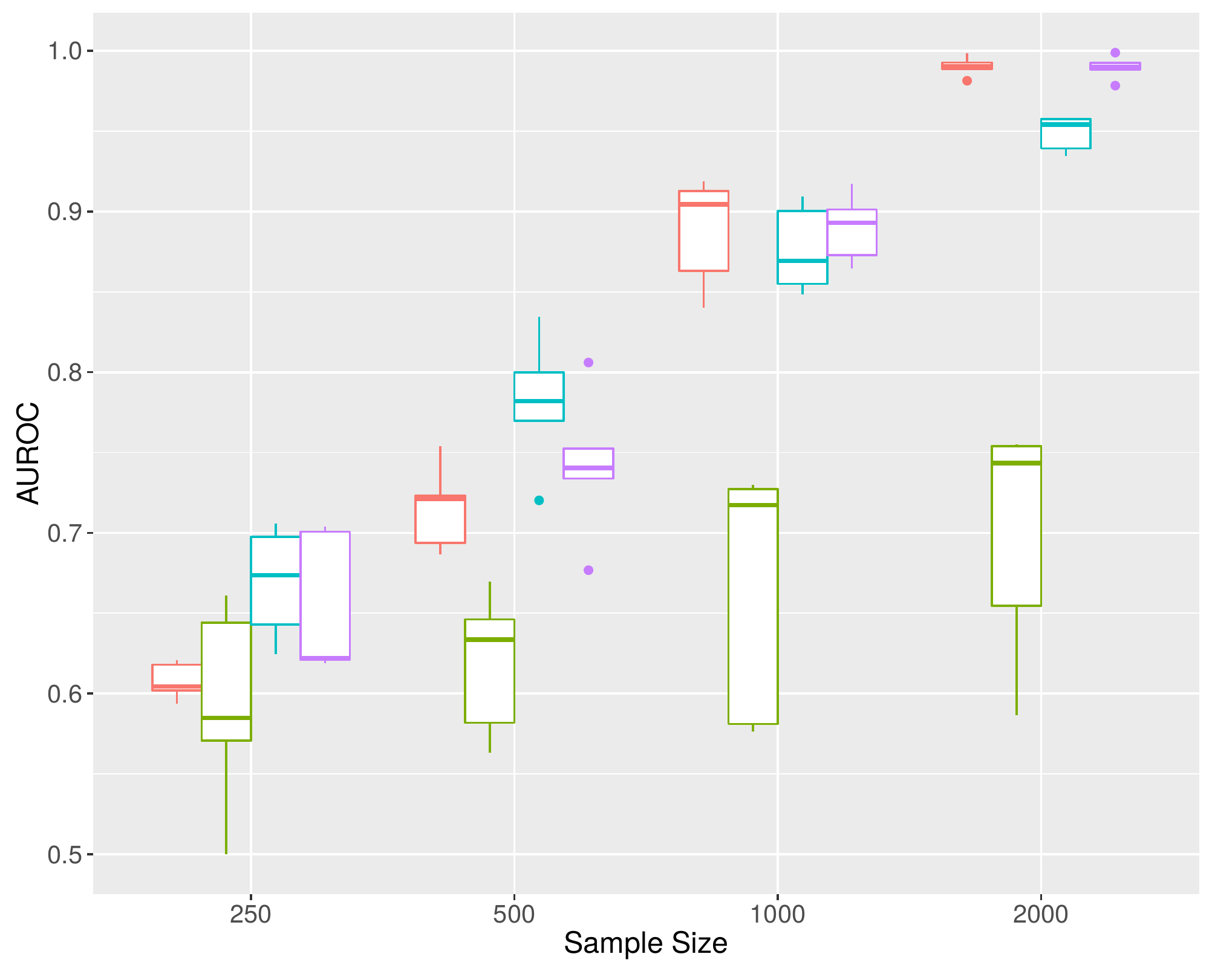}\\
  \includegraphics[width=0.45\textwidth]{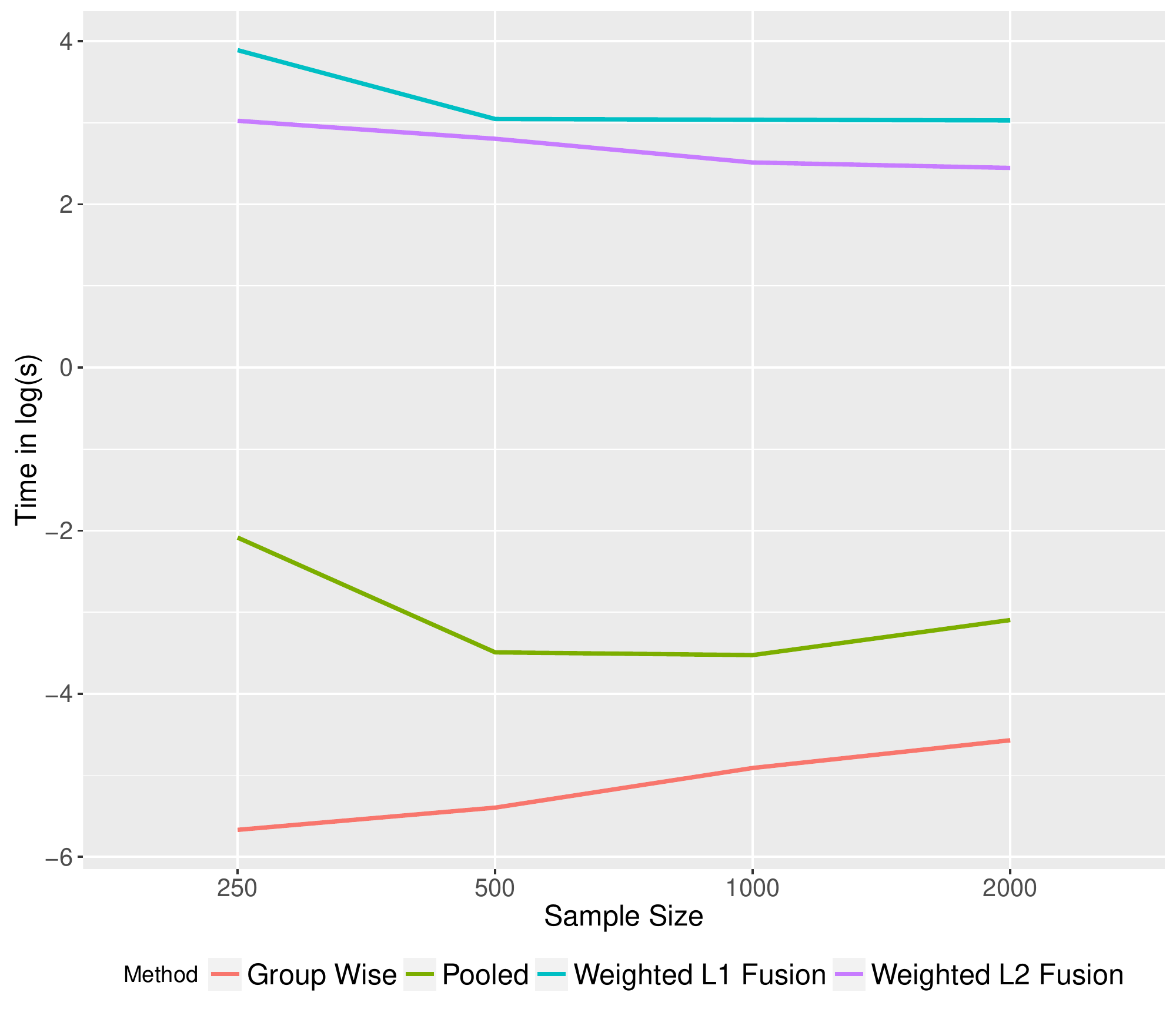}
  \caption{Simulated data, performance at varying sample sizes. Here, the number of subgroups is fixed at $K=9$ of which $K_0=4$ have shared coefficients in the underlying data-generating model (see text for details of simulation set-up).  Top left: Root mean squared error (RMSE; weighted by subgroup sizes). Top right: Area under the ROC curve (AUROC; with respect to the true sets of active variables with non-zero coefficients).    Bottom: Computational time taken in log seconds.}
  \label{fig:sim_vary_n}
\end{figure}

Figure \ref{fig:sim_vary_s} shows  performance when varying the number $K_0$
of subgroups with shared coefficients, with the total number of samples fixed at $n=250$.
Here, a smaller value of $K_0$ corresponds to less similarity between subgroup-specific coefficients in the underlying models.
At intermediate values of $K_0$ the fusion approaches offer gains over pooled and subgroup-wise analyses. This is because the pooled analyses are mis-specified due to the inhomogeneity of the data, while the subgroup-wise analyses, although correctly specified,
must confront limited sample sizes since they analyze each subgroup entirely separately.
In contrast, the fusion approaches are able to pool information across subgroups, but also  allow for subgroup-specific coefficients.
Importantly, even at the extremes of $K_0=1$ (separately drawn coefficients for each subgroup) and $K_0=9$ (all subgroups have exactly the same coefficients), the fusion approaches perform well.
This demonstrates their flexibility in adapting the degree of fusion.

Figure \ref{fig:sim_vary_n} shows  performance as a function of total sample size $n$. Here, the number of subgroups with identical coefficients is fixed at $K_0=4$. This gives a relatively weak opportunity for information sharing, since 5/9 groups have separately drawn  coefficients.
Since the true $\beta_k$'s are not identical, the pooled analysis is mis-specified and accordingly even at large sample sizes, it does not
catch up with  the other approaches. As expected, subgroup-wise analyses perform increasingly well at larger sample sizes. However, at smaller sample sizes the fusion approaches show some gains.

The  $\ell_1$ and $\ell_2$ fusion approaches seem similar in performance.
Our $\ell_2$ implementation
leverages the \texttt{glmnet} package and
is more computationally efficient than the $\ell_1$ approach.
For computational convenience, in examples below we show results from the $\ell_2$
approach only.

\section{Alzheimers disease: prediction of cognitive scores}
\label{sec_ADNI}
Here, we use data from the Alzheimer's Disease Neuroimaging Initiative (ADNI) \citep{mueller2005} to explore the ability of fusion approaches to estimate regression models linking clinical and genetic covariates to disease progression, as captured by cognitive test scores.

In 2014, ADNI made a subset of its data available for a DREAM challenge \citep{allen_16} and we use these data here. The dataset consists of a total of $n=767$ individuals who were followed up over at least 24 months. Cognitive function was evaluated using the mini-mental state examination (MMSE). At baseline, individuals were classified as either cognitively normal (CN), early mild cognitive impairment (EMCI), late mild cognitive impairment (LMCI) or diagnosed with Alzheimer's disease (AD). These form clinically-defined subgroups for our analysis. For the present analysis, we use only  genetic data (single nucleotide polymorphisms or SNPs) and clinical profile as covariates and disregard the neuroimaging data.

The task is to predict the slope of MMSE scores over a 24-month period. The total number of SNPs available is  $\sim10^7$. Filtering by linkage disequilibrium reduces this to  $\sim2 \times 10^6$. For
computational ease, we pre-selected 20,000 of this latter group
that gave the smallest residuals when regressed with the clinical variables against responses in the training set.
We note that this biases our analyses, but we emphasize that our goal in this section is not biomarker discovery but comparison between approaches all using the same (pre-selected) covariates.

\begin{figure}[htbp]
  \centering
  \includegraphics[width=0.55\textwidth]{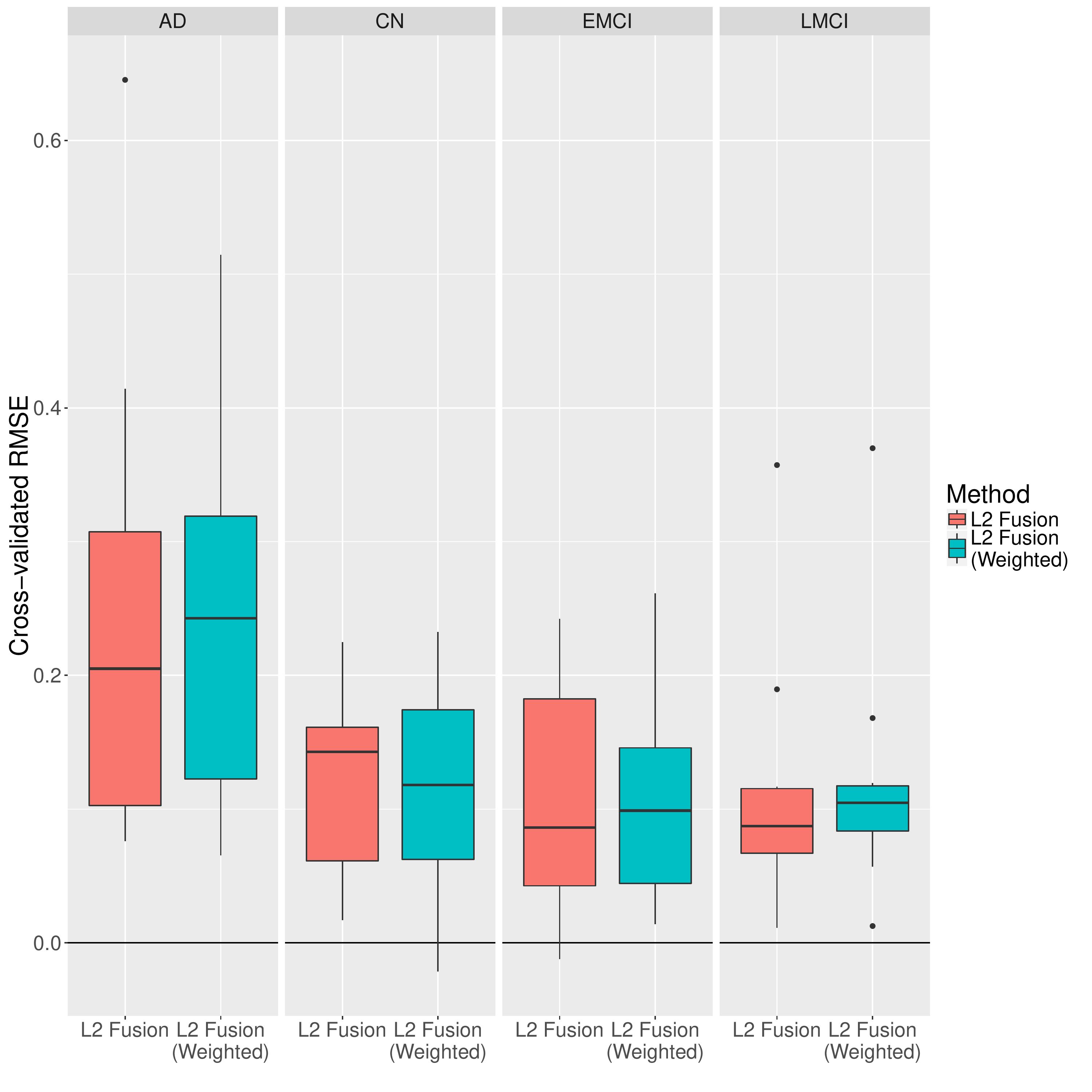}
  \caption{Alzheimers disease prediction results, ADNI data. Box plots showing difference in RMSE of fused methods compared to the pooled linear regression model (higher values indicate better performance by the fused methods). [Subgroup-wise analysis performed less well than pooled and is not shown; boxplots are over 10-fold cross-validation.]}
  \label{fig:adni_by_group}
\end{figure}

Figure \ref{fig:adni_by_group} shows root mean squared error (RMSE) separately for each of the four subgroups. The fusion approaches  offer substantial gains compared with pooled and subgroup-wise analyses (the latter performed very badly and are not shown in the figure). The biggest gain is for the AD subgroup. 
For the weighted fusion analysis the tuning parameters $\tau_{k,k'}$ were set using the distance between the means of each subgroup (in the space of genetic  and clinical variables). Weighting did not appear to improve performance.

\begin{figure}[htbp]
  \includegraphics[width=\textwidth]{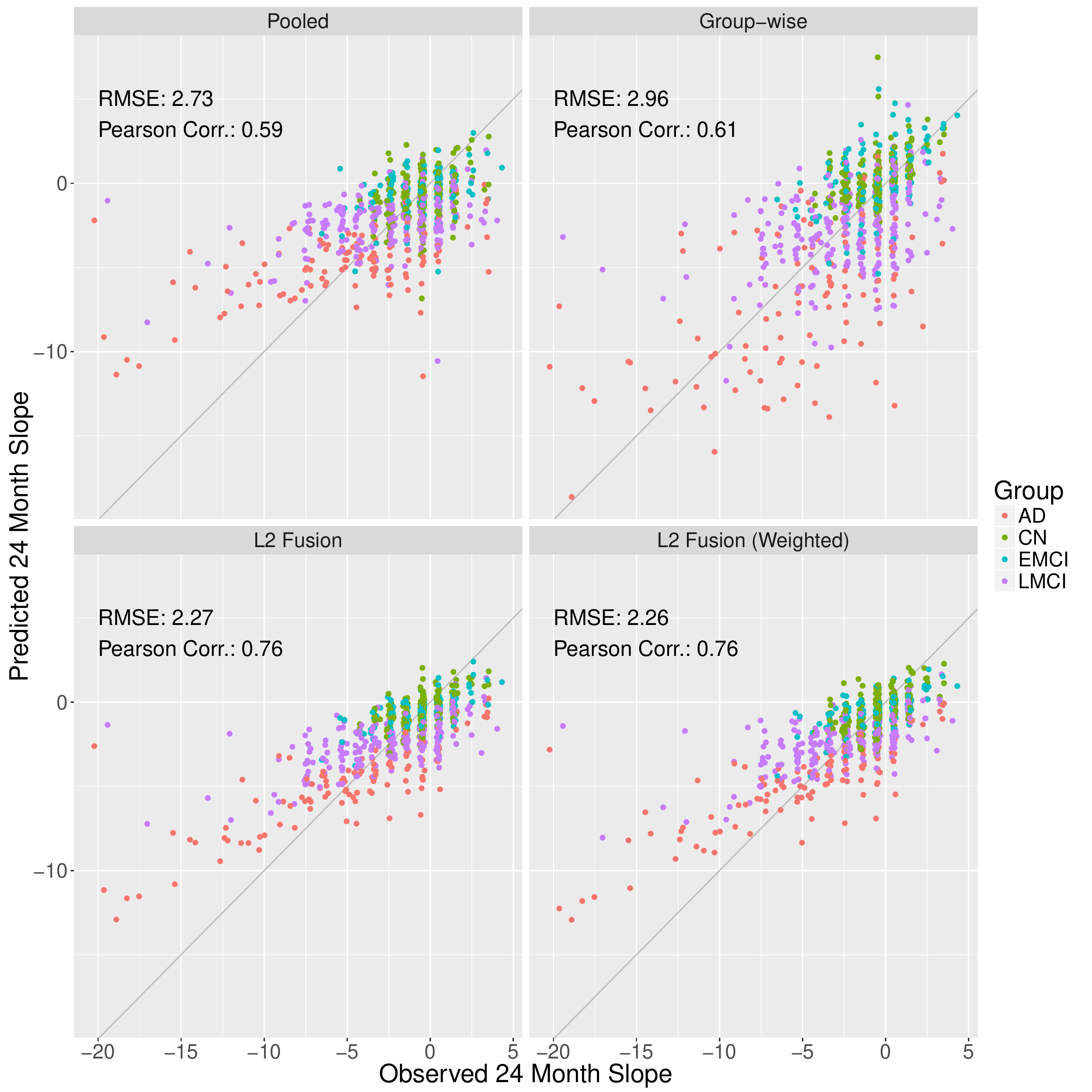}
  \caption{Alzheimers disease data, predicted vs. observed responses. Scatter plots show predicted and observed 24-month slopes for each of the standard and fused linear regression models. All predictions were obtained via 10-fold cross-validation.}
  \label{fig:adni_scatter}
\end{figure}

Figure \ref{fig:adni_scatter} shows scatter plots of predicted MMSE slopes versus the true slopes. The predictions shown were obtained in a held-out fashion via 10-fold cross-validation (CV), as were the RMSE and Pearson correlations shown.

\begin{figure}[htbp]
  \includegraphics[width=0.9\textwidth]{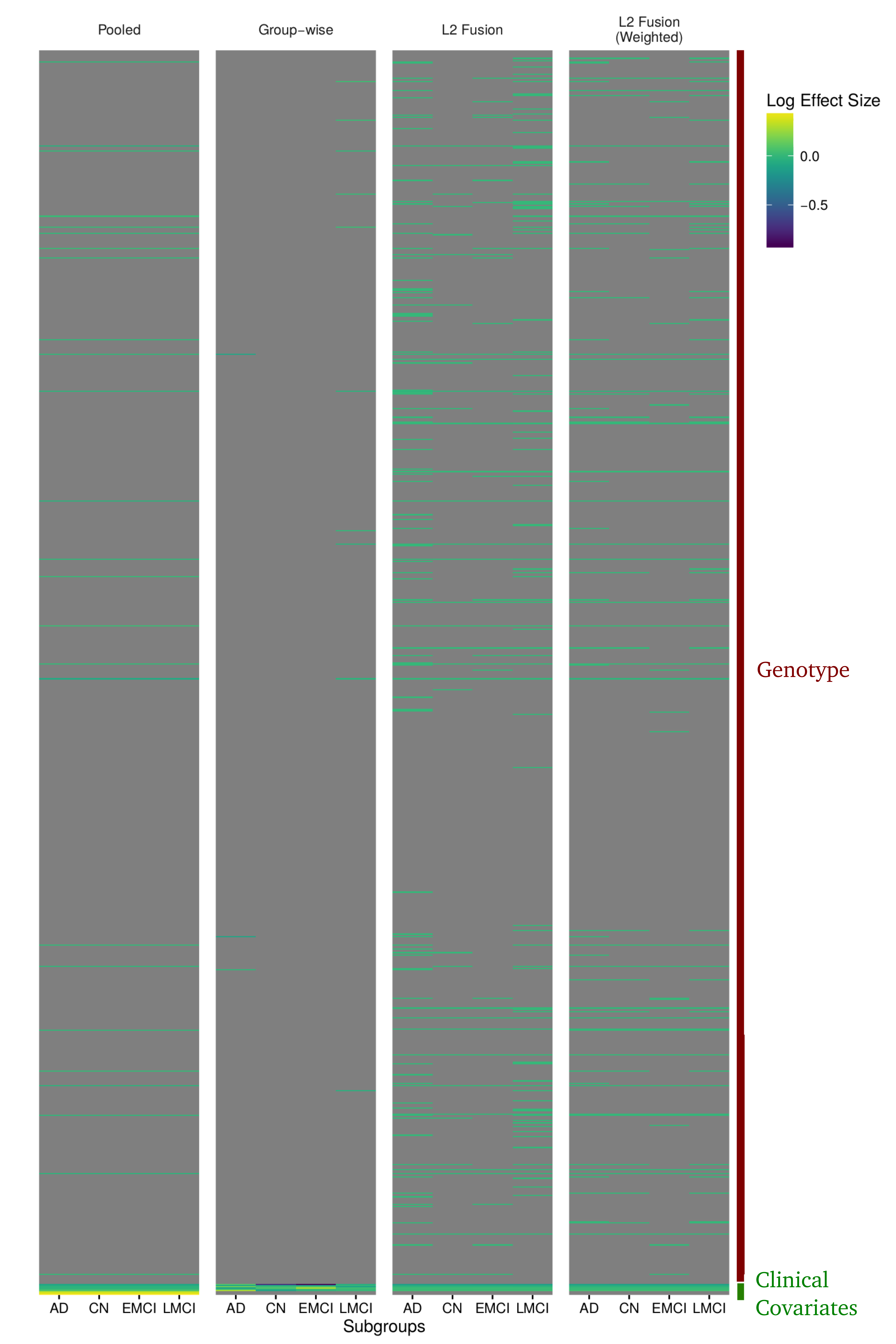}
  \caption{Alzheimers disease data, estimated regression coefficients. Heatmap showing estimated regression coefficients for the clinical variables and a representative subsample of the SNPs. Absolute coefficients are thresholded at $e^{-2}$ to improve readability.}
  \label{fig:adni_beta_comparison}
\end{figure}

Figure \ref{fig:adni_beta_comparison} shows a comparison of the estimated regression coefficients themselves. The subgroup-wise approach is much sparser than the other methods, likely due to the fact that it must operate entirely separately on each (relatively small-sample) subgroup. The pooled approach finds more influential variables but obviously there is no subgroup-specificity. The fused approach selects more variables than the subgroup-wise analysis, but there are many instances of subgroup-specificity in the estimates. 

\bigskip

\section{Prediction of therapeutic response in cancer cell lines}

The Cancer Cell Line Encyclopedia  \citep[CCLE, ][]{barretina2012} is a panel of 947  cancer cell lines
with associated molecular measurements and responses to 24 anti-cancer agents.
Here, we use these data to explore group-structured regression.
We treat the area above the dose-response curve as the response and use  expression levels of $\sim$ 20,000 human genes as covariates.
We treat the cancer types as subgroups $k$. After discarding cell lines with missing values, we arrive at  $n\sim 500$  samples.

\bigskip
\begin{figure}[htbp]
  \includegraphics[width=\textwidth]{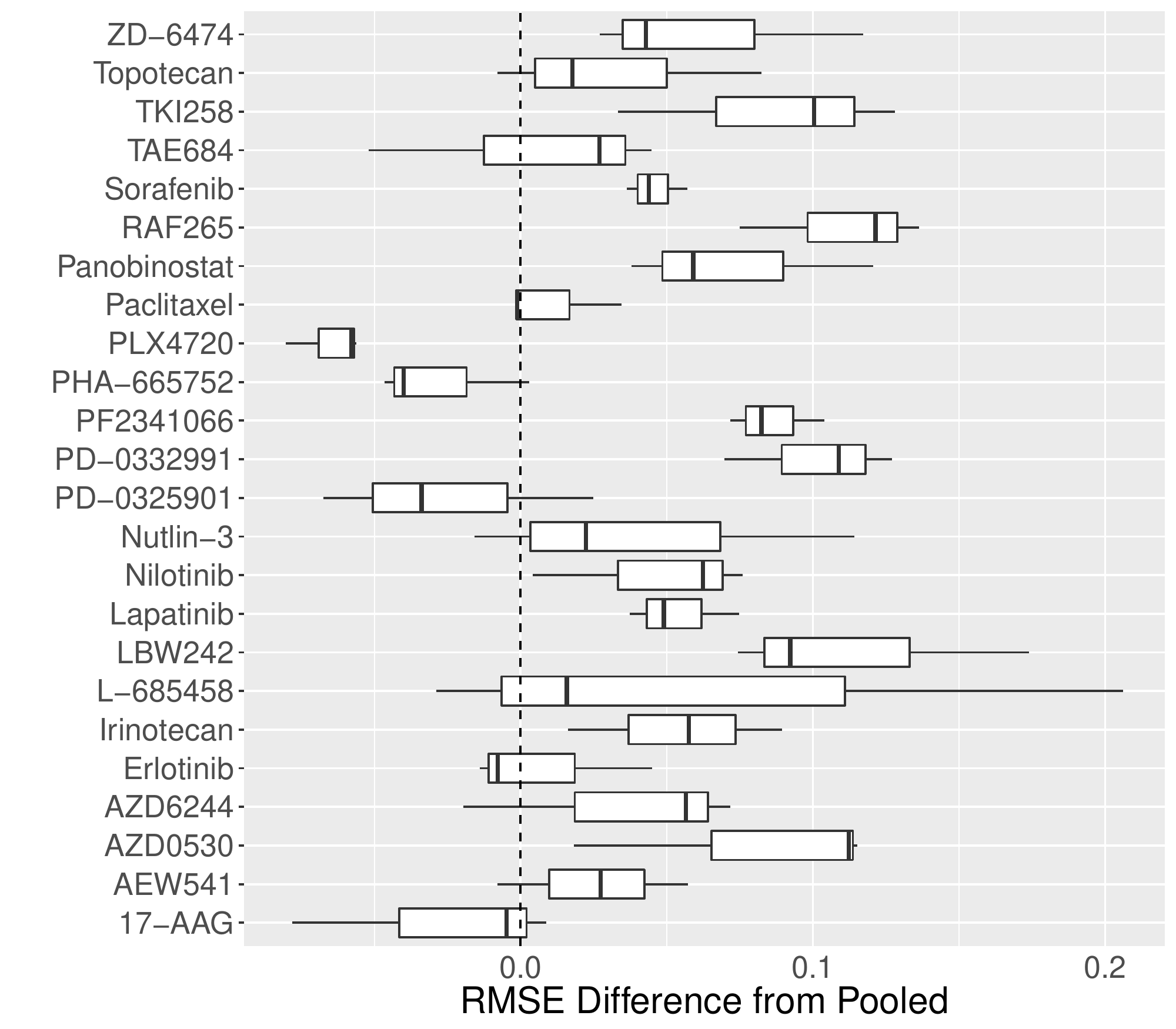}
  \caption{Cancer cell line therapeutic response prediction. Difference in weighted RMSE between L2 fusion approach and a pooled analysis. Results shown over 24 responses (anti-cancer agents) using data from the Cancer Cell Line Encyclopedia (CCLE); the dashed vertical line at zero indicates no difference, boxplots to the right indicate improvement (lower RMSE) over pooled.}
  \label{fig:ccle_results_all}
\end{figure}

Figure \ref{fig:ccle_results_all} shows results over all 24 responses (anti-cancer agents). We observe that for most responses the L2 fusion approach shows either improved or similar prediction performance to pooled in terms of RMSE (weighted by subgroup size). Weighted fusion shows a similar performance to unweighted fusion.

\begin{figure}[htbp]
  \includegraphics[width=0.49\textwidth]{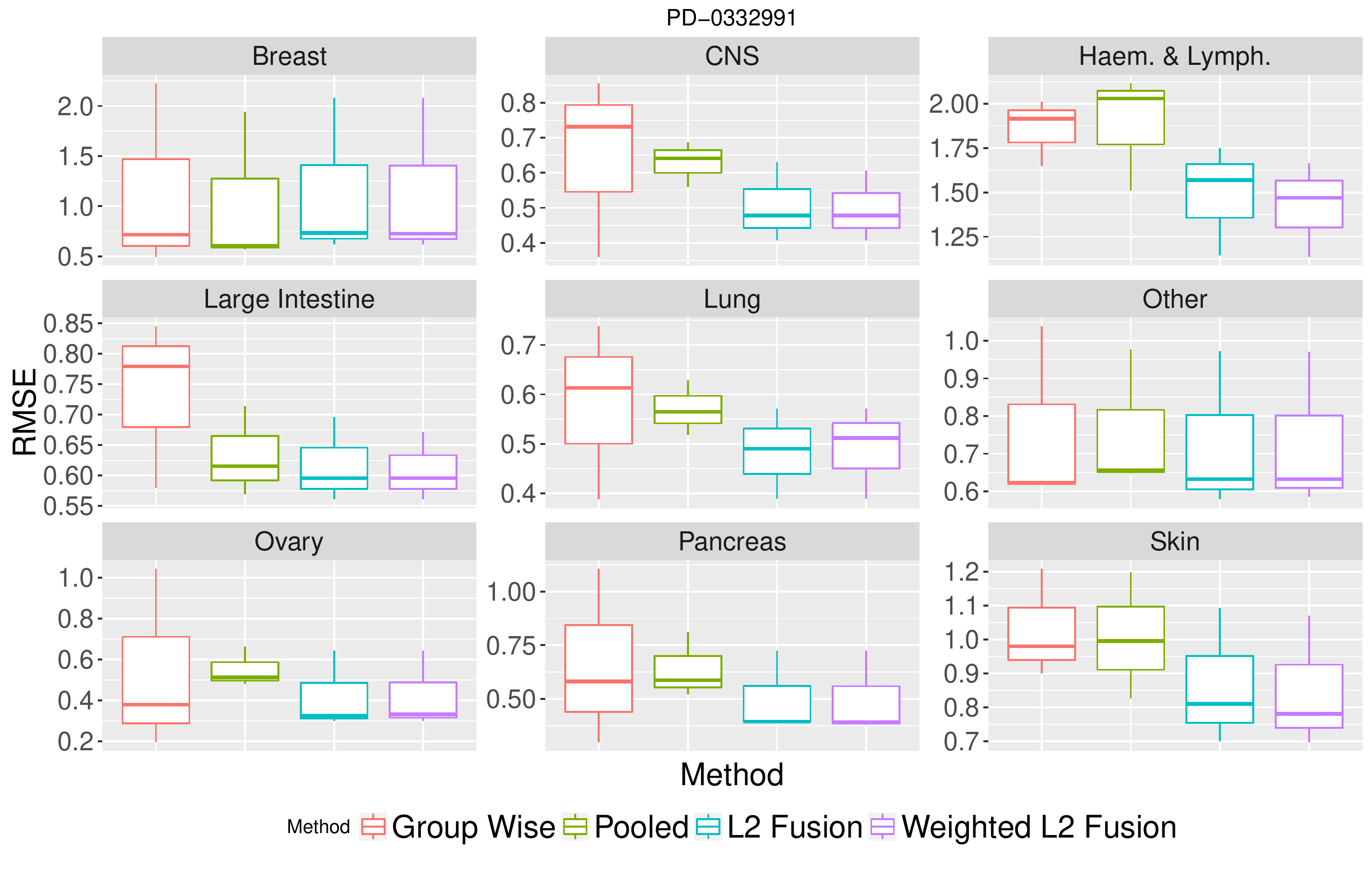}
  \includegraphics[width=0.49\textwidth]{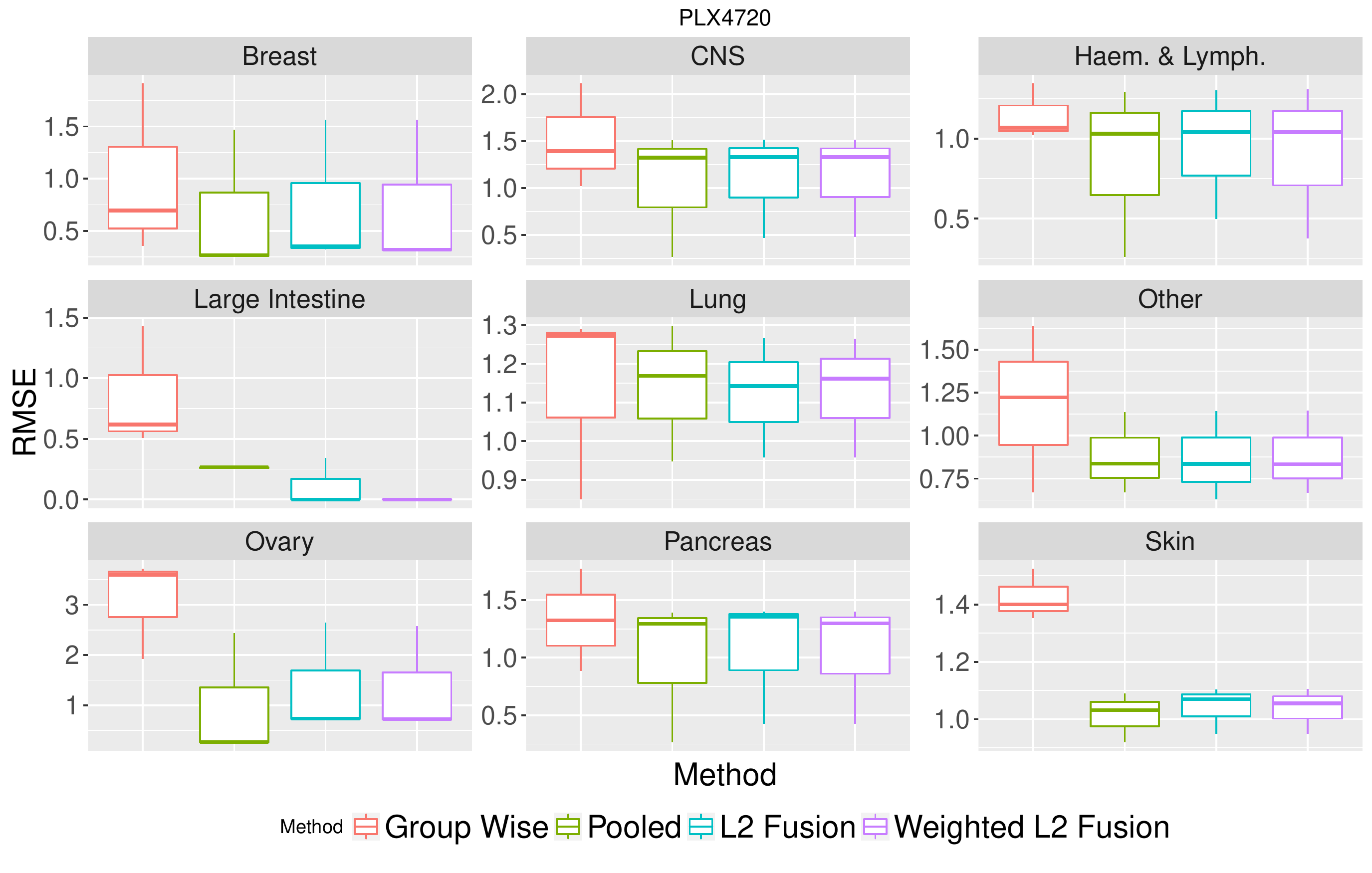}
  \caption{Cancer cell line therapeutic response prediction, broken down by subgroup (cancer type) for agents PD-0332991 and PLX4720.}
  \label{fig:ccle_results_drugs}
\end{figure}

Figure \ref{fig:ccle_results_drugs} shows results broken down by subgroup for two examples (responses PD-0332991 and PLX4720).
In the former case, the fusion approaches largely outperform pooled and subgroup-wise analyses. In the second, pooled is the best performer, although the fusion approaches are similar in most subgroups.

\section{ALS: prediction of  disease progression}
Amyotrophic lateral sclerosis (ALS) is an incurable neurodegenerative disease that can lead to death within three to four years of onset. However, about ten percent of patients survive more than 10 years. Prediction of disease progression remains an open question.
We use data from the PROACT database, specifically
data that were used in the 2015 DREAM ALS Stratification Prize4Life Challenge (data were retrieved from the PROACT database on 22/06/2015).
Our aim is not to optimize predictive performance {\it per se} but rather to provide a case study exploring the use of fusion approaches in a moderate-dimensional, clinical data setting.
In contrast to the Alzheimers example above, here the data are less high-dimensional and the subgrouping less clear cut (see below).

The data consist of observations from $n=2,393$ patients.
Each patient was enrolled in a clinical trial and followed up for a minimum of 12 months after the start of the trial. Disease progression is captured by a clinical scale called the ALS Functional Rating Scale (ALSFRS). The  task is to predict the slope of the ALSFRS score from 3 to 12 months (after the start of the trial). For each patient,
available covariates include ALSFRS scores for the 0-3 month period, demographic information and longitudinal measurements of clinical variables. We follow the featurization and imputation procedures devised by Mackay \citep[see][]{kuffner2015} and obtain a total of $p=615$ covariates.

Subgroups were defined as follows. The first subgroup consists of patients with disease onset  before the start of the trial.
The second subgroup consists of patients for whom onset was after the start of the trial and who have negative ALSFRS slope. The third subgroup of patients also had onset after the start of the trial but positive ALSFRS slope.
Thus, the subgroups reflect severity of onset.
 As we believe that the pre-trial onset group are likely to differ most from the others, we manually set the distance between groups 1 and the other two groups to 1 (the maximum), and set the distance between groups 2 and 3 to 0.1.

\begin{figure}[htbp]
  \centering
  \includegraphics[width=0.5\textwidth]{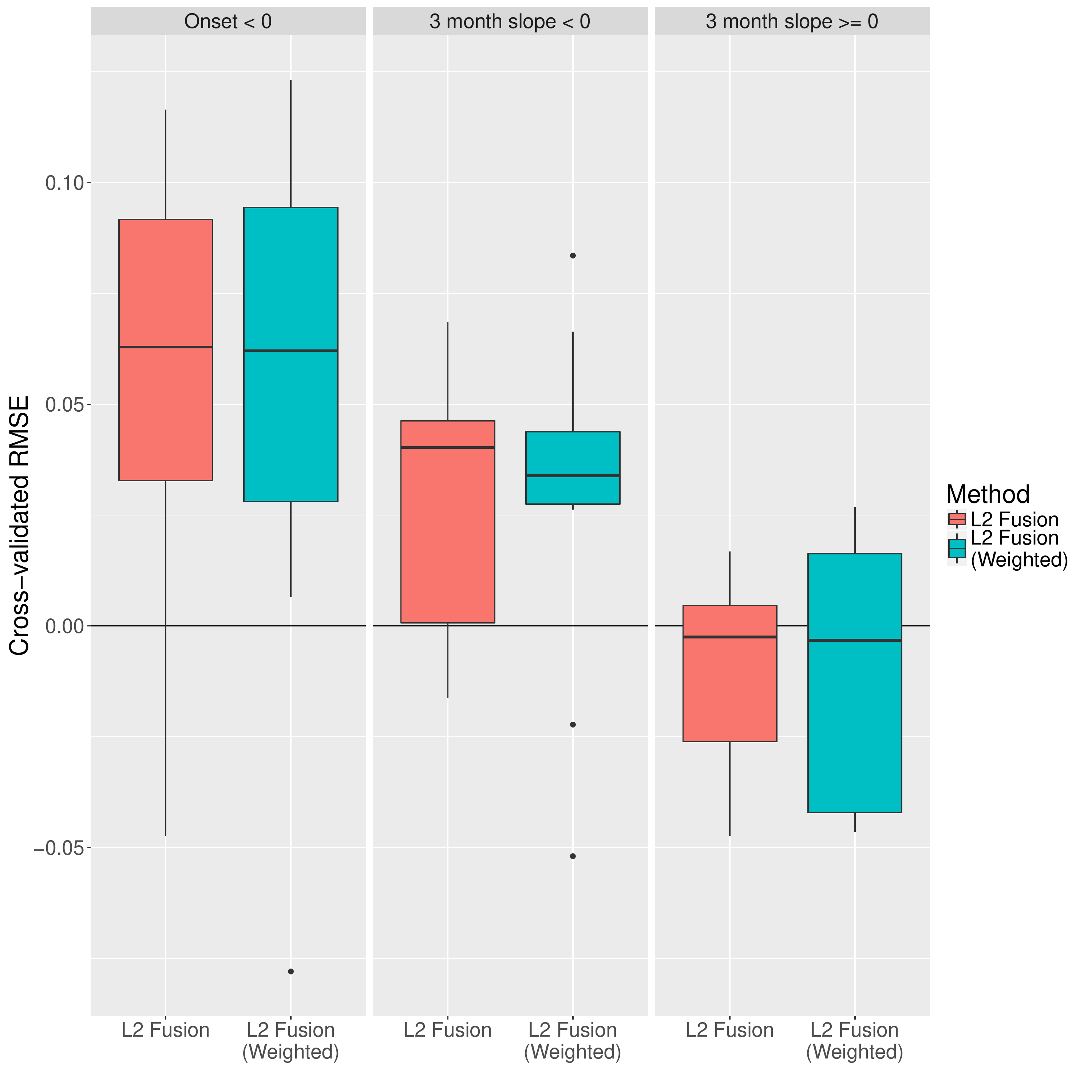}
  \caption{ALS prediction results. Box plots showing difference in RMSE of fused methods compared with the pooled linear regression model (higher values indicate better performance by the fused methods). [Subgroup-wise analysis performed less well than pooled and is not shown; boxplots are over 10-fold cross-validation.]}
  \label{fig:als_by_group}
\end{figure}

Figure \ref{fig:als_by_group}
shows (held-out) RMSEs by subgroup; we see that the largest improvement in prediction performance is in subgroup 1.
The fusion approach leads to a modest improvement. The difference between weighted and unweighted fusion is negligible\footnote{This dataset is different from the one reported in \citep{kuffner2015}, with larger variance in the slopes, and so RMSE values are not directly comparable; however, we note that performance for our methods compares favorably with that reported in the reference.}.


\bigskip


\section{Conclusions}
Many  biomedical datasets are heterogenous, spanning multiple disease types (or other biological contexts) that are related but  also expected to have specific underlying biology.
This means that large datasets are often usefully thought of as comprising several smaller datasets, that have similarities but that cannot be assumed to be identically distributed. Statistically efficient regression in these group-structured settings requires ways to pool information where useful to do so, whilst retaining the possibility of subgroup-specific parameters and sparsity patterns. We proposed a penalized likelihood approach for high-dimensional regression in the group-structured setting that provides group-specific estimates with global sparsity and that allows for information sharing between groups.

In any given application, even when there are good scientific reasons to suspect differences in regression models between subgroups, it is hard to know in advance whether the nature of any differences is such that a specific kind of joint estimation would be beneficial. For example, if sample sizes are small and groups  only slightly different, pooling may be more effective, or if the groups are entirely different, fusion of the kind we consider may not be useful.
This means that in practice, either simple pooling or subgroup-wise analysis may be   more effective than fusion. In our approach, the tuning parameter $\gamma$ (set by cross-validation) determines the extent of fusion in a  data-adaptive manner, and we saw in several examples that this appears successful in giving results that are  at worst close to the best of pooling and subgroup-wise analyses. For settings with widely divergent $n_k$'s, it may be important to allow tuning parameters to depend on $n_k$ (we did not do so) and to consider alternative formulations that allow for asymmetric fusion.




An appealing feature of our approach is that it allows for subgroup-specific sparsity patterns and parameter estimates that may themselves be of scientific interest. We discussed point estimation, but did not discuss uncertainty quantification for these subgroup-specific estimates. A number of recent papers have discussed significance testing for lasso-type models \citep[see e.g.][]{wasserman2009,lockhart2014,stadler2016} and we think some of these ideas could be used with the models proposed here.

\section{Software Availability}

The R code used for the experiments in this paper has been made available as R package \texttt{fuser} on GitHub: \url{https://github.com/FrankD/fuser}. Scripts for reproducing the results in this paper can be obtained at: \url{http://fhm-chicas-code.lancs.ac.uk/dondelin/SubgroupFusionPrediction}.

\section{Acknowledgements}

Data collection and sharing for the Alzheimer's data application was funded by  the  Alzheimer's  Disease Neuroimaging  Initiative  (ADNI)  (National  Institutes  of  Health  Grant  U01  AG024904)  and DOD  ADNI  (Department  of  Defense  award  number  W81XWH-12-2-0012). ADNI  is  funded by  the  National  Institute  on  Aging,  the  National  Institute  of  Biomedical  Imaging  and Bioengineering, and through generous contributions from the following: AbbVie, Alzheimer’s Association;  Alzheimer’s  Drug  Discovery  Foundation;  Araclon  Biotech;  BioClinica,  Inc.; Biogen;   Bristol-Myers   Squibb   Company;CereSpir,   Inc.;Cogstate;Eisai   Inc.;   Elan Pharmaceuticals,  Inc.;  Eli  Lilly  and  Company;  EuroImmun;  F.  Hoffmann-La  Roche  Ltd  and its  affiliated  company  Genentech, Inc.;  Fujirebio;  GE  Healthcare;  IXICO  Ltd.; Janssen Alzheimer    Immunotherapy    Research \&    Development,    LLC.;    Johnson \&    Johnson Pharmaceutical  Research \&  Development  LLC.; Lumosity; Lundbeck; Merck  \&  Co.,  Inc.; Meso  Scale  Diagnostics,  LLC.; NeuroRx  Research;  Neurotrack  Technologies;Novartis Pharmaceuticals Corporation; Pfizer Inc.; Piramal Imaging;Servier; Takeda Pharmaceutical Company;  and  Transition  Therapeutics. The Canadian  Institutes  of  Health  Research  is providing  funds  to  support  ADNI  clinical  sites  in  Canada.  Private  sector  contributions  are facilitated by the Foundation for the National Institutes of Health (www.fnih.org). The grantee organization is the Northern California Institute for Research and Education, and the study is coordinated by the Alzheimer’s Therapeutic Research Institute at the University of Southern 
California.  ADNI  data  are  disseminated  by  the  Laboratory  for  Neuro  Imaging  at  the University of Southern California.

\bibliography{sfl_theory}
\bibliographystyle{plainnat}


\end{document}